\documentclass[twocolumn,twoside,slac_two]{revtex4}
\usepackage{graphicx}
\usepackage{fancyhdr}
\begin{document}

\title{VERITAS Observations of Radio Galaxies}

%

\author{N. Galante, for the VERITAS Collaboration}
\affiliation{Harvard-Smithsonian Center for Astrophysics, 60 Garden Street, Cambridge, MA 02138, USA}

\begin{abstract}
Radio galaxies are the only non-blazar AGN detected in the VHE \mbox{($E > 100$ GeV)} band. These objects enable the investigation 
of the main substructures of the AGN, in particular the core, the jet and its interaction with the intergalactic environment. 
VERITAS observations have included exposures on a number of radio galaxies. Recently, the discovery by \emph{Fermi} of GeV 
emission from the radio galaxy NGC~1275 triggered VERITAS observations of this source. 
Results from the VERITAS observations of radio galaxies and future plans are presented.
\end{abstract}

\maketitle

\thispagestyle{fancy}


\section{INTRODUCTION}

The search for $\gamma$-rays from radio galaxies is  important for
the understanding of the dynamics and structure of jets in active galactic nuclei (AGN).
Even though radio galaxies are AGN with jets, their jet is not oriented toward the observer
and therefore the radiation produced by the jet is not Doppler-boosted towards
higher energies and luminosities, making them more challenging to detect in the 
very high energy (VHE: $E>100$~GeV) regime.
The discovery of VHE  $\gamma$-rays from the radio galaxy M~87 by the HEGRA
collaboration~\citep{Aharonian2003}, detected later by VERITAS~\citep{Acciari2008a},
and from NGC~5128 (Centaurus~A) by the HESS collaboration~\citep{Aharonian2009} has shown that
non-blazar AGN can produce very energetic photons from non-thermal processes.

Radio galaxies are classified into two main
families based on the morphology of their radio emission~\citep{FanaroffRiley},
whether it is core dominated (FR~I) or lobe dominated (FR~II),
with differences in the radio energetics and
in the discrete spectral properties~\citep{Zirbel1995}. The large number of features that
FR~I radio galaxies share with BL Lac type blazars suggests a possible unification between
the two sub-classes of AGN, in which FR I radio galaxies are BL Lac objects observed at larger jet viewing
angles~\citep{UrryPadovani}.

Evidence for synchrotron emission in radio to X-ray energies from both the extended structures and 
the core is well explained by relativistic particles moving in a beamed
relativistic jet~\citep{Ghisellini1993}. 
A commonly considered mechanism for HE-VHE (HE: high energy, 100~MeV$<E<$100~GeV) radiation is the synchrotron-self-Compton (SSC) 
process~\citep{Jones1974}, where the optical and UV synchrotron photons are up-scattered by the same 
relativistic electrons in the jet. Predictions concerning the inverse Compton (IC) component 
have long been established for the $\gamma$-ray 
emission~\citep{BloomMarscher1996} and frequency-dependent 
variability~\citep{Ghisellini1989}. Besides leptonic scenarios, several models also consider a hadronic origin for 
non-thermal emission in jets. Accelerated protons can initiate electromagnetic cascades or 
photomeson processes~\citep{Mannheim1993}, or directly emit synchrotron radiation \citep{Aharonian2002, Reimer2004}
and produce $\gamma$-rays through collisions with ambient gas \citep{Beall1999, Pohl2000}.

Modelling the blazar jet emission with a homogeneous SSC mechanism may imply particularly
high Lorentz factors, $\Gamma \gtrsim 50$, with consequent high Doppler factors and small beaming angles $\theta \simeq 1^\circ$
\citep{Kraw2002}. Such a small beaming angle is in conflict with the unification scheme according to which FR~I radio galaxies
and BL~Lac objects are the same kind of object observed at different viewing angles. Moreover,
these high values for the Doppler factor are in disagreement with the small apparent velocities observed
in the sub-parsec region of the TeV BL Lac objects Mrk~421 and Mrk~501 \citep{Marscher1999}.
These considerations suggest a more complicated geometry, for example 
a decelerating flow in the jet with a consequent gradient in the Lorentz
factor of the accelerated particles and a smaller average $\Gamma$ \citep{Georganopoulos2003}.
As a result of this gradient, the fast upstream particles interact with the downstream 
seed photons with an amplified energy density, because of the Doppler boost due to the relative Lorentz factor
$\Gamma_\mathrm{rel}$. The IC process then requires less extreme values for the
Lorentz factor and allows larger values for the beaming angle.
In a similar way, a jet spine-sheath structure consisting of a faster internal spine 
surrounded by a slower layer has been also suggested for the broadband non-thermal emission of VHE BL Lac 
objects~\citep{Ghisellini2005}. An inhomogeneous jet with a slow component may explain the HE-VHE emission observed in radio galaxies at larger angles              
($\theta_\mathrm{layer} = 1/\Gamma_\mathrm{layer} \sim 20^\circ$). 
Observation of the VHE component from radio galaxies
is therefore significant for the AGN jet modeling. In this work an overview of the observations
of radio galaxies by VERITAS is presented.

\section{OBSERVATIONS}

\subsection{NGC~1275}

NGC~1275 (Perseus~A, 3C~84) is a nearby \mbox{($z = 0.018$)} radio galaxy located in the center of the Perseus cluster. 
It is one of the most unusual early-type galaxies in the nearby Universe. Its radio emission is core dominated, but it also has strong
emission lines. In addition, the emission line system shows a double structure, corresponding to both a high-velocity and a low-velocity 
system. The puzzling nature of NGC~1275 makes it difficult to definitively classify it in a standard AGN sub-class.
It has been recently detected in high energy $\gamma$-rays by \emph{Fermi}~\cite{Abdo2009}, with a flux between 100~MeV and 25~GeV described by a power 
law with photon index $-2.17$. 

\begin{figure*}[t]
\centering
\includegraphics[width=135mm]{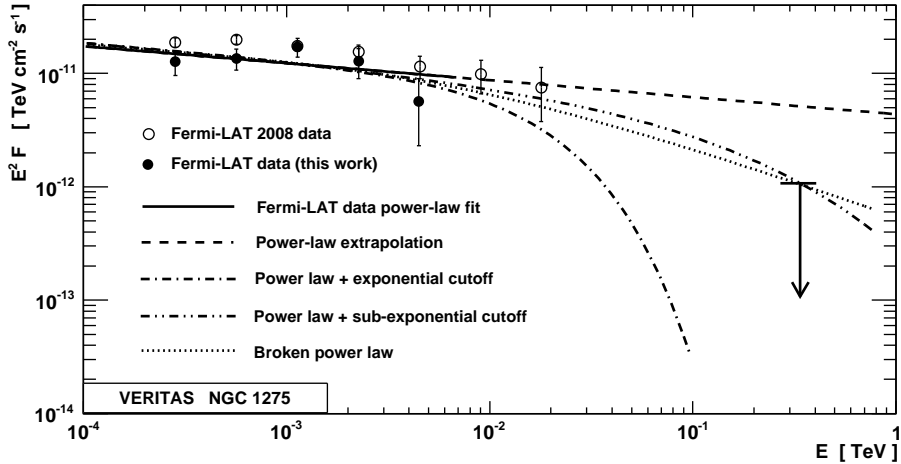}
\caption{NGC~1275 spectrum and the VERITAS upper limit on the differential flux at 
the decorrelation energy 338~GeV (standard cuts). The solid circles with error bars are the measurement
by the \emph{Fermi} $\gamma$-ray space telescope during the VERITAS observation campaign. 
Empty circles with error bars are the measurement presented in~\cite{Abdo2009} from the energy-binned analysis. 
The solid line is the power-law fit to the \emph{Fermi} data. 
The dashed line is the extrapolation of the power-law. The dotted-dashed line is the fit of
a power law with an exponential cutoff at 18~GeV. The double-dotted dashed line
is the fit of a power law with a sub-exponential cutoff at 120~GeV and the dotted line is
the smooth broken power law fit of a break energy at 16~GeV. All fits are done on the \emph{Fermi} data
analyzed in this work (solid circles).} \label{fig:ngc1275}
\end{figure*}

VERITAS observed the source between January and February 2009 for a total amount of good-quality data of 7.8~hours. Additional 
\emph{Fermi}-LAT data simultaneous to the VERITAS observational campaign have been analyzed, reporting a lower flux by a factor of 1.37 and a similar photon 
index $-2.15$ compared to the 2008 published \emph{Fermi}-LAT spectrum. A differential upper limit at the decorrelation energy of ~340~GeV is calculated and is incompatible \mbox{($P(\chi^2)=3.6\times10^{-4}$)} with an extrapolation of the \emph{Fermi} measured spectrum (fig.~\ref{fig:ngc1275}). A deviation from the power-law regime is therefore a likely explanation. Three possible models have been considered: a power law with an exponential cutoff, with a sub-exponential cutoff and a broken power law, all rising from different electron energy distributions in the jet as the absorption by the extra-galactic background light (EBL) is excluded due to the proximity of the galaxy.
The three possibilities are equally compatible \mbox{($P(\chi^2)=0.2$)} with the VERITAS upper limit.
The estimated cutoff energies are $E_\mathrm{exp}\approx20$~GeV and $E_\mathrm{subexp}\approx 120$~GeV in the case of an exponential and sub-exponential
cutoff respectively, and $E_\mathrm{b}\approx 16$~GeV in the case of a broken power law. For details of the analysis see~\cite{Acciari2009b}.

The result of the observation is rather interesting. It shows, for the first time, that there can be a deviation from the power-law regime
in a radio-galaxy spectrum at an energy of the order of $~100$~GeV or lower. This is the first example of a scientific
result obtained by VERITAS in conjunction with \emph{Fermi}.

\subsection{3C~111}

\begin{figure*}[t]
\centering
\includegraphics[width=70mm]{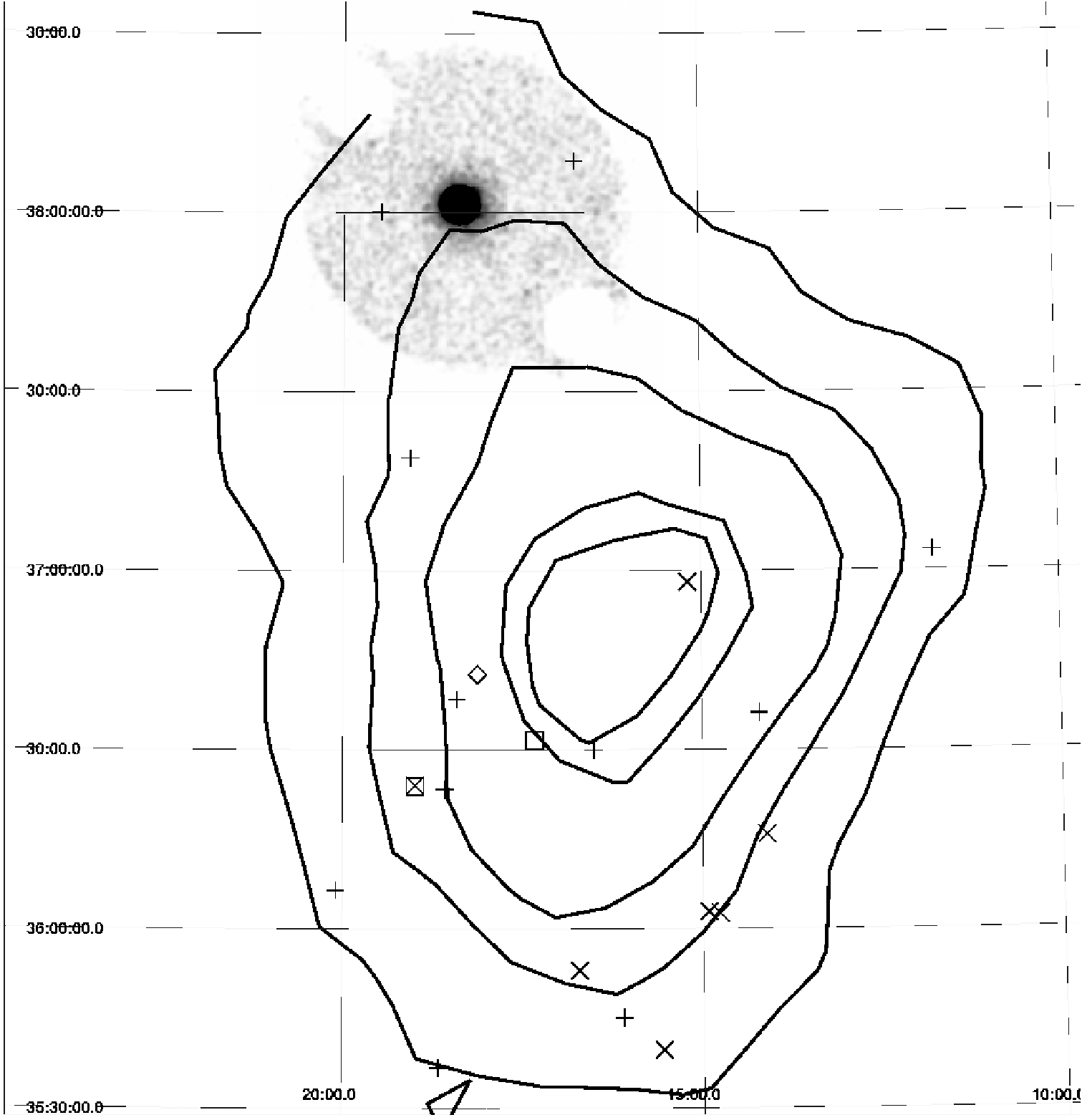}
\includegraphics[width=70mm]{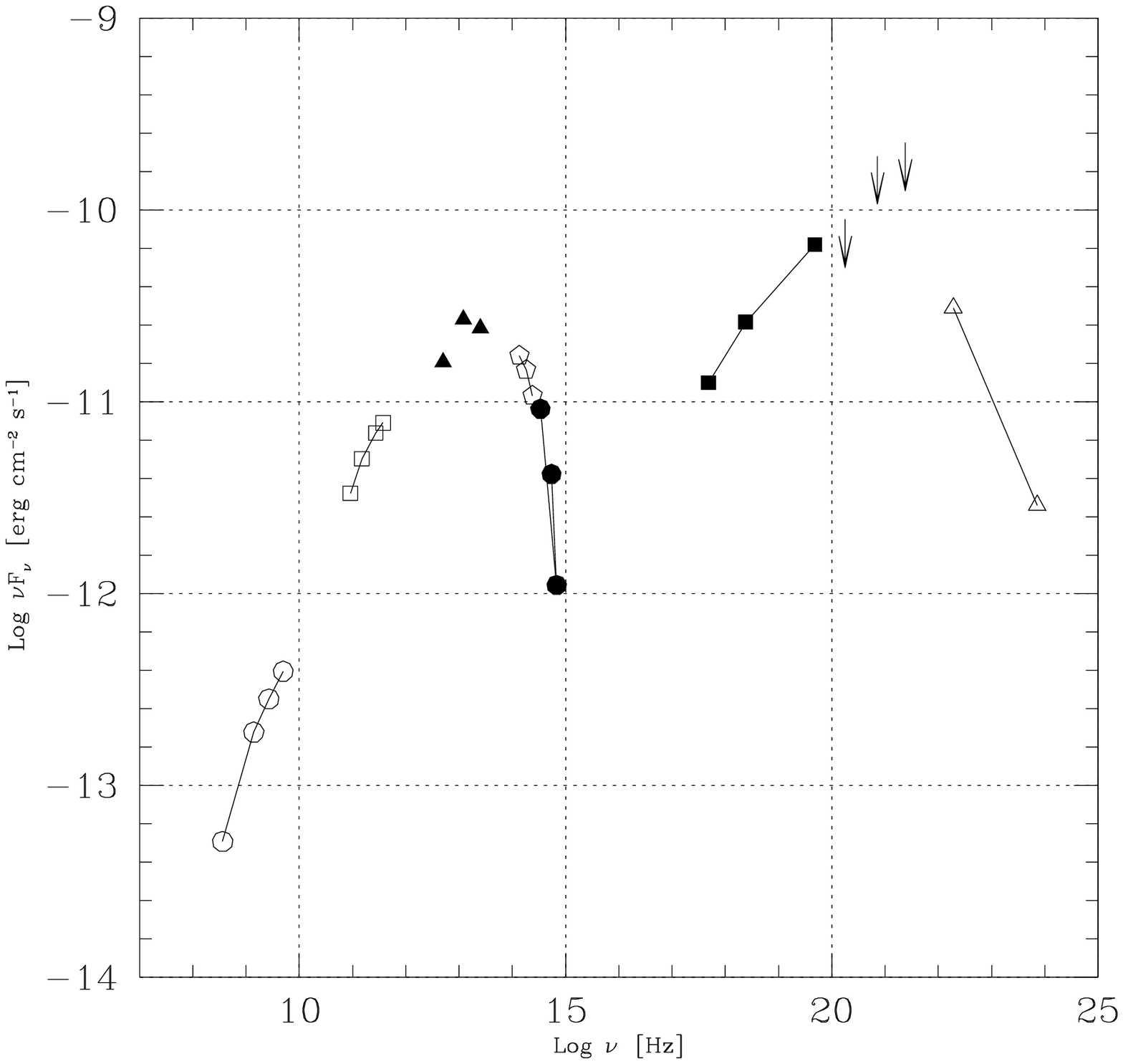}
\caption{(\emph{left}) The BeppoSAX-MECS (2-10~keV) image superimposed on the EGRET $\gamma$-ray probability contours at 50\%, 68\%,
95\%, 99\% and 99.9\% confidence level. Crosses: ROSAT faint sources; diamonds: ESS sources; plusses: NVSS radio sources;
squares: GBT radio sources. (\emph{right}) Broadband SED of 3C~111. Open circles: radio; open squares: mm-band;
filled triangles: IRAS; filled circles: optical; open pentagons: infra red (2MASS); filled squares: BeppoSAX; open triangles: EGRET; arrows:
$2\,\sigma$ upper limits by COMPTEL. Both figures from~\cite{Sguera2005}.} \label{fig:3c111}
\end{figure*}

3C~111 is a near ($z=0.0485$) FR-II radio source whose central component is coincident with a broad-line Seyfert~1 galaxy.
The radio morphology shows a double-lobe/single-jet structure with the jet emerging at an angle of $\sim63^\circ$~\cite{Linfield1984}.
The central component is variable on time scales of a few months. Hints of superluminal behaviour are observed~\cite{Preuss1988, Preuss1990}.
The radio spectrum is flat with an index of $-0.3$ between 6~cm and 80~cm~\cite{Becker1991}.
Strong emission is detected in the mm and infra-red bands too~\cite{Bloom1994, Golombeck1988}. In the X-ray band, 3C~111 has been detected by
many instruments with a long-term variability within a factor of 5~\cite{Reynolds1998}.
The broadband spectral energy distribution (SED) shows a double-peaked structure (see fig.~\ref{fig:3c111} right) with typical blazar-like
features and the source has been suggested to be a misaligned blazar~\cite{Sguera2005}.

The radio galaxy 3C~111 has been suggested as a counterpart for the unidentified EGRET $\gamma$-ray source
3EG~J0416+3650~\cite{Hartman1999}. This is a $\gamma$-ray source located at $l=162^\circ.2 \quad b=9^\circ.97$,
i.e. close to the galactic plane, with a 95\% confidence error radius of 38'.2. However, since the optical position of the
radio galaxy is outside the 99\% confidence level contour of the EGRET $\gamma$-ray source ($\sim 76$~arcmin separation),
the probability of the association between the two sources is rather low ($P=0.019$)~\cite{Mattox2001}.

However, additional hint supporting the association between 3C~111 and 3EG~J0416+3650 can be found in~\cite{Sguera2005}. 
Due to the large uncertainty on the EGRET $\gamma$-ray source position,
12~X-ray and radio sources can be found inside the $3\, \sigma$ confinement error box~(see fig.~\ref{fig:3c111} left). Nevertheless, the radio
galaxy 3C~111 is among the 12 sources the only object that is known to emit both in radio and X-rays, with the hardest and
strongest X-ray flux.

The $5.3\,\sigma$ detection reported by EGRET with an average flux above 100~MeV of \mbox{$1.3\times 10^{-7}$ cm$^{-2}$ s$^{-1}$}
with a simple power-law photon index $-2.59$ makes this $\gamma$-ray source interesting for an instrument
like VERITAS. If detected, given its higher angular resolution, VERITAS would be able to definitely identify the $\gamma$-ray emitter
with the underlying object.

VERITAS observed the radio galaxy 3C~111 during fall 2008 at a zenith angle range between $15^\circ$ and $30^\circ$. All data taken
under bad weather conditions or with technical problems have been discharged. Finally, a total
amount of about 11~hours has remained for analysis purposes. No VHE signal has been detected,
a 99\% confidence level upper limit above the analysis threshold of 300~GeV has been derived.
The result is reported in table~\ref{tab:1}.

\subsection{M~87}

M~87 is a radio galaxy located in the Virgo cluster at a distance of 16~Mpc~\cite{Macri1999}. 
Originally detected at TeV energies at 4 sigma significance by HEGRA~\cite{Aharonian2003}
and finally detected over 5 sigma by HESS~\cite{Aharonian2006}, it has been later detected also by VERITAS~\cite{Acciari2008a}. The substructures of the jet are
well studied in the X-ray, optical and radio wavelengths~\cite{Wilson2002}, with an estimated angle of $20^\circ - 40^\circ$ toward the line of sight.
Its proximity and spatially-resolved structures at all wavelengths make M~87 a unique laboratory to study the jet physics, especially
for the related mechanisms to the VHE $\gamma$-rays production. Given its peculiarity, an extensive VERITAS-led coordinated multi-wavelength observational
campaign, involving all major imaging air Cherenkov telescopes (IACT) currently operating, VERITAS, MAGIC and HESS,
and othe X-ray and radio instruments, namely Chandra and VLBA, has been performed in 2008. 
Correlation studies of this broad-band observational campaign resulted in the identification of the region responsible for the origin
of the $\gamma$-ray emission~\cite{Acciari2009b}.
A dedicated contribution has been therefore presented at this Symposium~\cite{Wagner2009}.

\section{CONCLUSIONS}

VERITAS observed three radio galaxies during the last two years. Only in one case, the already-known $\gamma$-ray emitter M~87,
the observation resulted in a VHE $\gamma$-ray emission detection. Given the peculiarity of the radio galaxy M~87, that makes it
a unique laboratory for the study of the blazar astrophysics, in particular for the jet-related processes, VERITAS coordinated
an observational campaign together with the major IACT partners and other X-ray and radio partners. The broad-band observational campaign
resulted in the discovery of the region responsible of the $\gamma$-ray emission.

The observation of other two radio galaxies, 3C~111 and NGC~1275, did not result in a VHE detection. However,
a joint work together with \emph{Fermi}-LAT on NGC~1275 resulted in the identification of a variation from the power-law
regime at an energy of the order of $\approx 100$~GeV or lower, a previously unknown feature of radio-galaxies
$\gamma$-ray component.

\begin{table}[t]
\begin{center}
\caption{VERITAS upper limits on the observed radio galaxies VHE flux. The five columns represent:
the source name; the period of observation; the energy threshold for that specific analysis;
the total observation time of good quality data; the 99\% confidence level integral flux upper limit
in cm$^{-2}$~s$^{-1}$.}
\begin{tabular}{|l|c|c|c|c|}
\hline \textbf{Source} & \textbf{Obs. Period} & \textbf{$E_\mathrm{th}$} &
\textbf{$T_\mathrm{obs}$} & \textbf{Flux U.L.}
\\
\hline \textbf{3C~111} & 10/08 - 12/08 & 300~GeV & 11~hr & $3.5\times 10^{-12}$\\
\hline \textbf{NGC~1275} & 01/09 - 02/09 & 190~GeV & 8~hr & $5.11\times 10^{-12}$ \\
\hline \textbf{M~87} & \multicolumn{4}{|c|}{See~\cite{Wagner2009}}\\
\hline
\end{tabular}
\label{tab:1}
\end{center}
\end{table}

\bigskip 
\begin{acknowledgments}
This research is supported by grants from the US Department of Energy, the US National Science Foundation, and the Smithsonian Institution, by NSERC in Canada, by Science Foundation Ireland, and by STFC in the UK. We acknowledge the excellent work of the technical support staff at the FLWO and the collaborating institutions in the construction and operation of the instrument. 
\end{acknowledgments}

\bigskip 

\end{document}